\documentclass[aps,twocolumn,amsmath,amssymb,superscriptaddress]{revtex4-2}

\pdfoutput=1

\usepackage[T1]{fontenc}

\usepackage{graphicx}

\usepackage[colorlinks=true,citecolor=blue,linkcolor=red,urlcolor=blue]{hyperref}
\usepackage[figure,figure*]{hypcap}

\let\Im\relax
\DeclareMathOperator\Im{Im}

\let\s\textsubscript

\makeatletter
\newcommand\cdw{\protect\@dblarg\@cdw} 
\newcommand\@cdw[2][]{#1\,\texttimes\,#2} 
\makeatother

\def\Zagreb{Centre for Advanced Laser Techniques, Institute of Physics, 10000~Zagreb, Croatia}

\def\Bremen{U~Bremen Excellence Chair, Bremen Center for Computational Materials Science, and MAPEX Center for Materials and Processes, Universität Bremen, 28359~Bremen, Germany}

\def\Louvain{European Theoretical Spectroscopy Facility, Institute of Condensed Matter and Nanosciences, Universit\'e catholique de Louvain, 1348~Louvain-la-Neuve, Belgium}

\def\Wavre{WEL Research Institute, 1300~Wavre, Belgium}

\begin{document}

\sloppy

\title{Understanding the origin of superconducting dome in electron-doped MoS\s2 monolayer}

\author{Nina Girotto Erhardt}
\email{ngirotto@ifs.hr}
\affiliation\Zagreb

\author{Jan Berges}
\affiliation\Bremen

\author{Samuel Ponc\'e}
\affiliation\Louvain
\affiliation\Wavre

\author{Dino Novko}
\email{dino.novko@gmail.com}
\affiliation\Zagreb

\begin{abstract}
We investigate the superconducting properties of molybdenum disulphide (MoS\s2) monolayer across a broad doping range, successfully recreating the so far unresolved superconducting dome.
Our first-principles findings reveal several dynamically stable phases across the doping-dependent phase diagram.
We observe a doping-induced increase in the superconducting transition temperature $T_c$, followed by a reduction in $T_c$ due to the formation of charge density waves (CDWs), polaronic distortions, and structural transition from the H to the 1T$'$ phase.
Our work reconciles various experimental observations of CDWs in MoS\s2 with its doping-dependent superconducting dome structure, which occurs due to the \cdw1 H to \cdw2 CDW phase transition.
\end{abstract}

\maketitle

\section*{Introduction}

Novel two-dimensional (2D) materials exhibit unparalleled and highly tailorable physical and chemical properties, making them appealing for both fundamental and technological reasons~\cite{Fiori2014,Kumbhakar2023}.
With a rich diversity of phases, including superconducting (SC)~\cite{Ye2012,Morosan2006,Ludbrook2015,Ichinokura2016,Nagamatsu2001,Kortus2001,Sajadi2018}, ferroelectric~\cite{Wang2022,Zheng2020}, (anti)ferromagnetic~\cite{Xu2022,Huang2017}, charge density wave (CDW)~\cite{BinSubhan2021,Rossnagel2011}, topological insulator~\cite{Li2019}, and excitonic insulator~\cite{Jia2022}, all reachable by tuning temperature, pressure or doping, it is of great interest to understand the phase diagrams of 2D materials.

Molybdenum disulphide (MoS\s2) is a quasi-2D layered semiconducting transition metal dichalcogenide (TMD) well known for its exceptional optical properties and strong excitonic features~\cite{Mak2010,Wang2012}.
When doped with excess electrons, MoS\s2 exhibits a complex, yet largely unexplored phase diagram, including various structural phases and many-body features such as multi-valley superconductivity~\cite{Ye2012,Ge2013,Costanzo2016}, Holstein polarons~\cite{Kang2018,GarciaGoiricelaya2019,vanEfferen2023,vanEfferen2024}, and CDWs~\cite{BinSubhan2021,Rosner2014,Piatti2018,Zhuang2017}.
The ground state of the MoS\s2 monolayer is the \cdw1 H phase, with a characteristic graphene-like hexagonal structure.
Mechanically and chemically exfoliated MoS\s2 samples differ in structure.
While the former results in the \cdw1 H structure, in the latter multiple phases coexist~\cite{Eda2011,Xiao2019,Zhao2018,Jin2018,Lin2014,Cheng2014}, such as 1T and 1T$'$ phases with various periodicities~\cite{Zhuang2017,Wypych1998}, mostly due to doping-induced structural modifications.
Recently, a CDW phase with \cdw2 and \cdw{$2 \sqrt 3$} reconstructions was shown to exist in potassium-intercalated MoS\s2~\cite{BinSubhan2021}.
Intriguingly, the SC transition temperature $T_c$ in electron-doped MoS\s2 follows a characteristic dome structure~\cite{Ye2012,Costanzo2018}, resembling the phase diagrams of other unconventional correlated materials such as high-$T_c$ copper oxides~\cite{Keimer2015}, iron-based pnictides and chalcogenides~\cite{Paglione2010}, SrTiO\s3~\cite{Koonce1967}, 1T-TiSe\s2~\cite{Morosan2006}, and Weyl semimetal WTe\s2~\cite{Pan2015} or MoTe$_2$~\cite{Paudyal2020}.
Even though the SC mechanism in MoS\s2 is well studied for lower carrier concentrations and it is understood in terms of multi-valley phonon-mediated pairing~\cite{Ge2013,Piatti2018}, the origin of the dome structure is still unresolved.
Along with the various coexisting many-body interactions, the presence of polarons, CDWs, and the H to 1T$'$ structural phase transition all inherently change the structure of MoS\s2, such that reconstructing a doping-dependent SC dome in MoS\s2 is theoretically challenging.
For the aforesaid systems hosting a SC dome, several mechanisms leading to a dome formation have been discussed in the literature, including the doping-induced Lifshitz transition in the Fermi surface~\cite{Ge2013}, the parametrization of McMillan's formula for $T_c$~\cite{Rosner2014}, anharmonic damping close to the phase transitions~\cite{Setty2022}, soft-phonon-mediated pairing~\cite{Appel1969,Edge2015,Yang2020,Ma2021}, and dynamical electron-electron interaction~\cite{Das2015}.
Interestingly, the impact of a doping-induced change of structure on $T_c$ has not yet been investigated in MoS\s2, even though CDW and SC phases seem to coexist.
It is not clear whether CDW supports or suppresses a SC phase~\cite{Joe2014,Kusmartseva2009,Johannes2008,Yu2021,Rossnagel2011,Zhu2022}, but both phases employ the same conduction electrons.
As the CDW phase is suppressed, the electronic density of states (DOS) at the Fermi level increases.
Consequently, the SC phase is able to receive more electrons for a SC condensate.
In general, this results in a simultaneous reduction of a CDW $T_c$ and increase of a SC $T_c$~\cite{Morosan2006,Morris1975,Gabovich2001,Nagata1992}, and it could provide a valid explanation for the SC dome in the CDW systems such as MoS\s2.

Theoretical endeavors, which include electron-phonon coupling (EPC) calculations of SC $T_c$ in the \cdw1 H-MoS\s2 phase within the isotropic Migdal-Eliashberg formalism~\cite{Eliashberg1960}, McMillan's formula~\cite{McMillan1968}, or the Allen-Dynes formula~\cite{Allen1975}, lead to large EPC constants and $T_c$'s~\cite{Ge2013,Rosner2014,Fu2017}.
These results could be improved with including anharmonic effects, which would lower the $T_c$ and delay the onset of CDW~\cite{Setty2024}, while the anisotropic Migdal-Eliashberg formalism might be important for MoS\s2 and low-dimensional systems~\cite{Margine2013}.
These effects were recently taken into account, while the doping mechanism was included in a field-effect configuration~\cite{Marini2022}, which resulted in the accurate prediction of a SC $T_c$, and a suppression of the EPC constant by an order of magnitude.
However, in that study, the SC dome was not found and the existence of CDW in the experimental doping range was ruled out.

Here we conduct a comprehensive first-principles investigation of the doping dependent phase diagram of MoS\s2 and find various stable phases across a large doping range.
Structural changes of MoS\s2 lead to a dome structure of the SC $T_c$ and encompass the experimental findings of a SC dome and CDW phase.
We perform nonadiabatic EPC calculations~\cite{Girotto2023,Berges2023,Novko2020}, and find that it leads to a considerable reduction compared to the adiabatic EPC constants and $T_c$.
Moreover, we find that $T_c$ can be further lowered by taking spin-orbit coupling (SOC) into account, which is known to determine the possible pairing symmetries~\cite{Yuan2014} and to enhance the upper critical magnetic field~\cite{Lu2015,Zhou2016,Zheliuk2019}.
With this study, we successfully fill in the missing gap in comprehending the complicated phase diagram of MoS\s2.
We show that the $T_c$ increases for the original H phase and reaches a maximum value in a region where a normal H phase and \cdw2 CDW coexist.
For even larger doping concentrations, the 1T$'$ phase and \cdw2 CDW ordering are stabilized with $T_c$ being lowered, which in total forms a characteristic dome structure as observed in the experiments~\cite{Ye2012,Costanzo2018}.
Our model calculations based on the tight-binding approximation (TBA)~\cite{Schobert2024} confirm the dome structure of $T_c$, and additionally reveal localized polaronic distortions for carrier concentrations in the phase diagram where neither the original \cdw1 H structure nor the \cdw2 CDW phase are stable.

\begin{figure}[!t]
    \centering
    \includegraphics[width=\columnwidth]{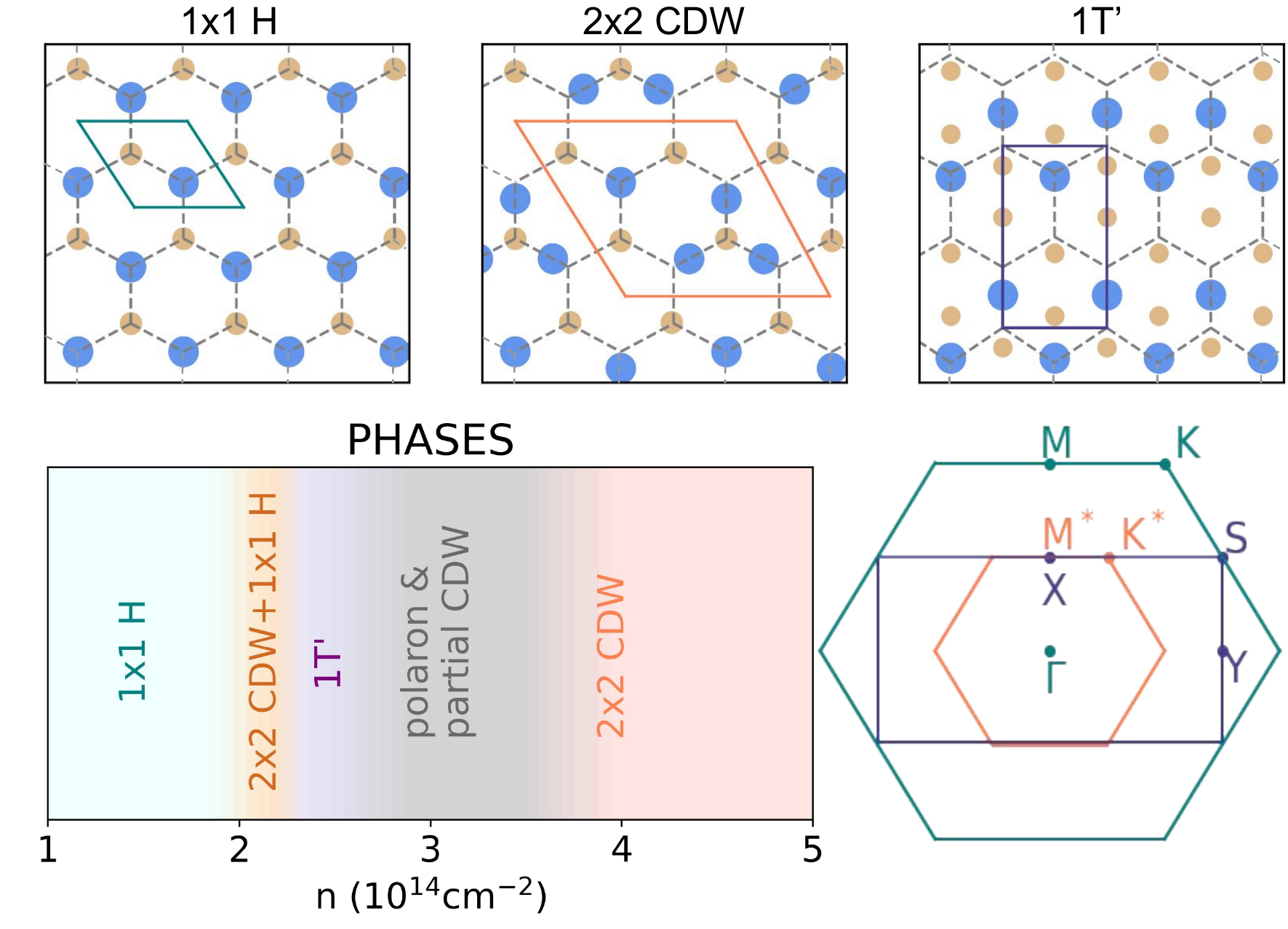}
    \caption{\textbf{Phase diagram of electron-doped MoS\s2 with various dynamically stable phases.}
    In the upper row, we show the three distinct phases appearing in the phase diagram together with their primitive cells.
    For visualization, the atomic structures are shown with exaggerated atomic displacements (10 times) and on top of a hexagonal mesh.
    Each of the phases from the first row is stable in some doping range, as color-coded in the phase diagram.
    For the grey shaded area our TBA model predicts the stabilization of polaronic distortions and partial CDWs.
    The corresponding Brillouin zones are displayed in the second row with the same color code from the phase diagram.}
    \label{fig:fig1}
\end{figure}

\section*{Results}

MoS\s2 can be doped by gating or intercalation reaching large charge densities.
In Fig.~\ref{fig:fig1} we show a doping-dependent phase diagram of dynamically stable structures that MoS\s2 assumes, as obtained from density functional theory (DFT), including the \cdw1 H phase, \cdw2 CDW structure with distorted atoms, and 1T$'$ phase.
Between the regions where the 1T$'$ and high-doping \cdw2 CDW phases are stable, our TBA simulations reveal polaronic distortions and partial CDW coverage.
Our aim is to calculate the EPC strengths and $T_c$ across these phase transitions.

\begin{figure*}[!t]
    \centering
    \includegraphics[width=0.8\textwidth]{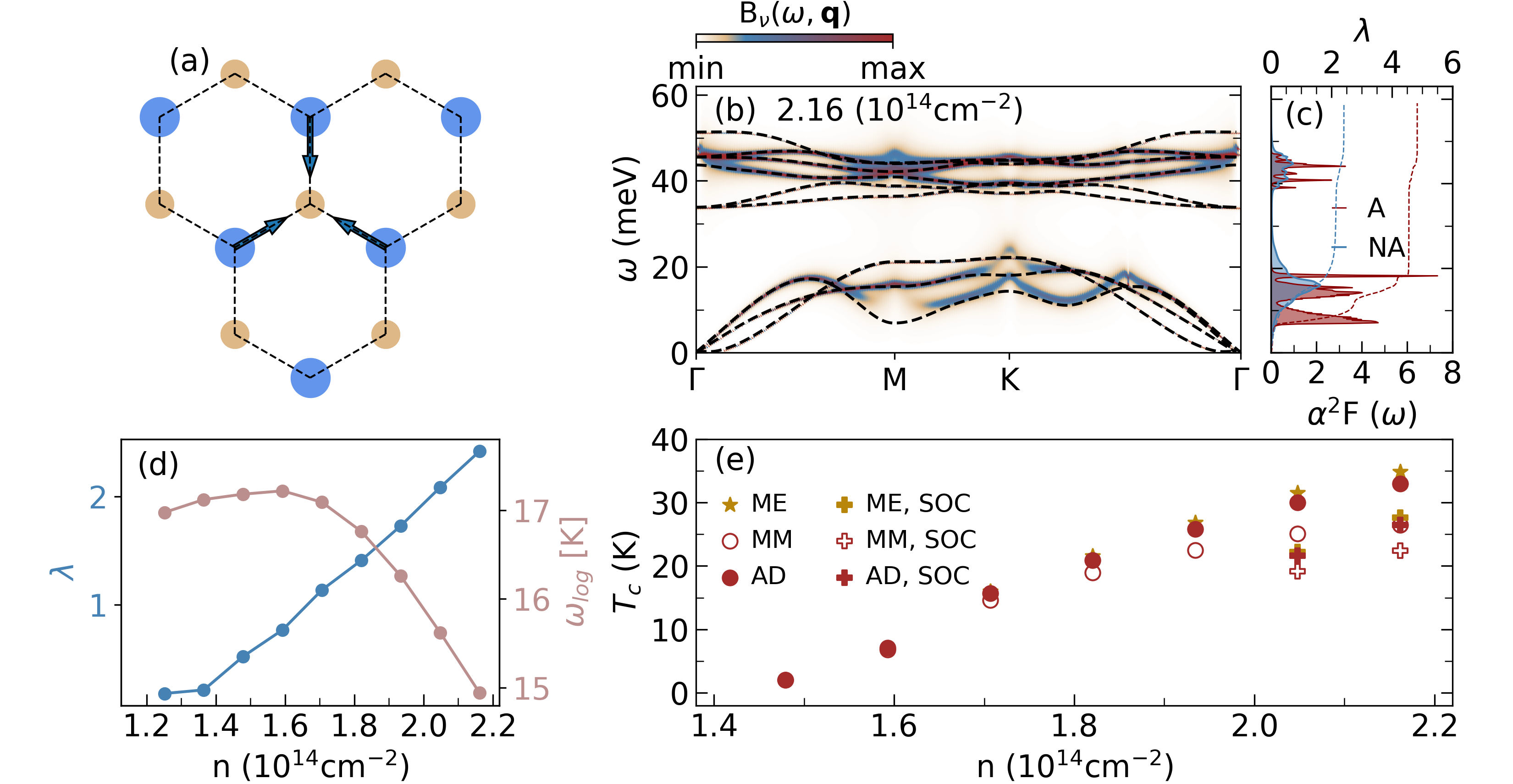}
    \caption{\textbf{Electron-phonon properties of MoS\s2 in the \cdw1 H phase.}
    (a)~Schematics of the linear combination of the strongly-coupled lowest-energy acoustic phonon eigendisplacements of A\s1 symmetry at all $\textbf{q}=M$ points, see Supplementary Figure~3.
    (b)~Phonon spectral function of MoS\s2 at 800\,K, with the adiabatic phonon band structure denoted with dashed black line.
    (c)~Eliashberg spectral function in the adiabatic (A, red) and nonadiabatic (NA, blue) approximations together with the corresponding EPC cumulative constant $\lambda(\omega)$.
    (d)~EPC strength $\lambda$ and logarithmic average frequency $\omega_{\log}$ computed with NA phonons.
    (e)~Superconducting $T_c$ calculated using McMillan's (MM) formula (hollow red circles), the Allen-Dynes (AD) formula (full red dots), and the isotropic Migdal-Eliashberg (ME) formalism (golden stars).
    Cross symbols denote the calculations which include spin-orbit coupling.}
    \label{fig:fig2}
\end{figure*}

\textbf{The low-doping \cdw1 H phase.}
Our calculated phonon dispersions indicate a dynamically stable \cdw1 H structure up to the excess charge concentration of $2.38 \times 10^{14}\,\mathrm{cm}^{-2}$, when the acoustic phonon of A\s1 symmetry at the M point of the Brillouin zone (BZ), shown in Fig.~\ref{fig:fig2}(a), induces the instability.
This critical doping $n_c$ is sensitive to the choice of the electronic temperature in the DFT electronic structure calculations, and varies from $n_c = 1.5 \times 10^{14}\,\mathrm{cm}^{-2}$ to $n_c = 2.4 \times 10^{14}\,\mathrm{cm}^{-2}$ when $T_{\rm scf}$ is between 100\,K and 800\,K, as one can see in Supplementary Figure~1.
Here we use $T_{\rm scf}=800$\,K since the choice of lower electronic temperature would require denser momentum grids.
This strongly-coupled acoustical phonon describes the motion of Mo atoms towards the S atom and by increasing doping, its frequency softens.
Its strong coupling to electrons manifests itself as a large broadening in the phonon spectral function $B_{\nu}(\mathbf{q},\omega)$ [see Fig.~\ref{fig:fig2}(b)] and as a large peak in the Eliashberg spectral function, $\alpha^2F(\omega)$ [see Fig.~\ref{fig:fig2}(c)].
Here we also show how the $\alpha^2F(\omega)$ and the EPC constant $\lambda$ in a nonadiabatic (NA) formalism (see Methods) differ from their adiabatic (A) counterparts~\cite{Girotto2023}.
With the largest contribution coming from the A\s1 soft mode, the NA frequency renormalization and broadening effects strongly renormalize the EPC properties.
In fact, the NA effects reduce the EPC strength $\lambda$ by half, which mostly comes from the blueshifted Kohn anomalies at the M and K points of the BZ.
In the second row of Fig.~\ref{fig:fig2}, we show doping-dependent quantities calculated from the NA phonon spectrum.

The doping-dependent EPC constant $\lambda$ and the corresponding $\omega_{\log}$ for the \cdw1 H stable part of the phase diagram in Fig.~\ref{fig:fig2}(d) form a dome, but we rule out the possibility that the SC dome arises due to $\omega_{\log}$ in the $T_c$ formula~\cite{Rosner2014}.
In fact, we find that the $T_c$ is monotonically increasing according to McMillan's formula, the Allen-Dynes formula, as well as the isotropic Migdal-Eliashberg formalism in agreement with Ref.~\cite{Marini2022} [see Fig.~\ref{fig:fig2}(e)].
For the two largest dopings we show how the inclusion of the SOC lowers the $T_c$ by approximately 30\%.
An increase of the Coulomb pseudopotential $\mu^*$ can also reduce the value of $T_c$ by a few percent (see Supplementary Figure~2).
Note that our computed $T_c$ values overestimate the experiments, and that more quantitative results would be obtained with the additional inclusion of anharmonicity and proper gating geometry~\cite{Marini2022}.

\begin{figure}[!t]
    \centering
    \includegraphics[width=0.95\columnwidth]{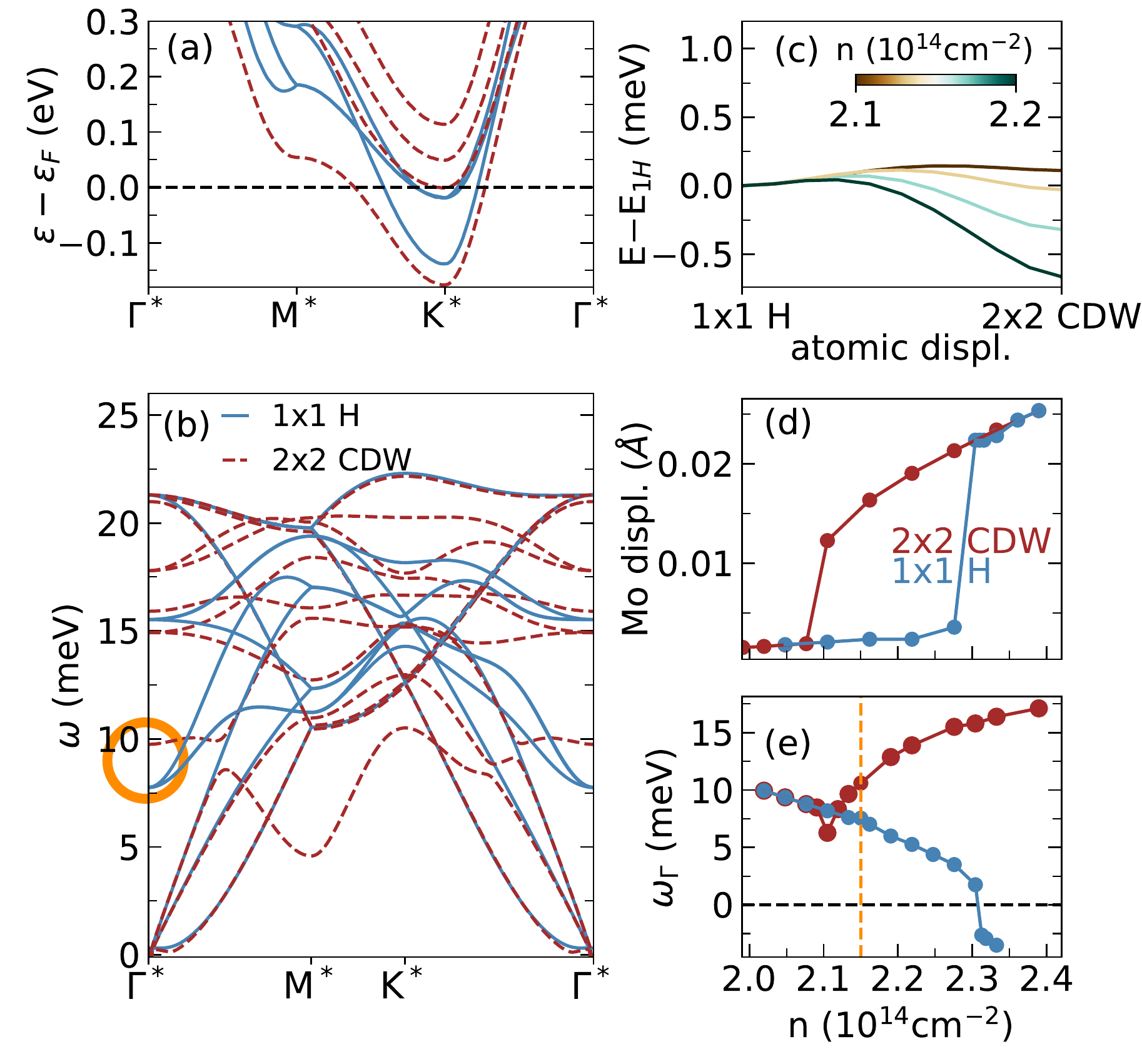}
    \caption{\textbf{Coexistence of the \cdw1 H and \cdw2 CDW phases and corresponding electron and phonon properties}.
    (a)~Electron band structures comparison of the two phases for $n = 2.15 \times 10^{14}\,\mathrm{cm}^{-2}$.
    (b)~Phonon dispersion comparison, with the soft mode from (e) denoted with an orange circle.
    (c)~Potential energy surface between the two phases for a range of different dopings.
    (d)~Hysteresis plot of the doping-dependent displacement of the Mo atoms.
    (e)~Doping-dependent frequency of the $\Gamma^*$ mode, backfolded from the M point of the \cdw1 unit cell.
    The orange vertical line corresponds to $n = 2.15 \times 10^{14}\,\mathrm{cm}^{-2}$.
    See Fig.~\ref{fig:fig1} for the definition of the starred ($^*$) high-symmetry points.
    }
    \label{fig:fig3}
\end{figure}

\textbf{Coexistence of \cdw2 CDW and \cdw1 H phases.}
Slightly before the instability, we find that a \cdw1 H structure coexists with a \cdw2 CDW phase.
In analogy to the 1T-TiSe\s2~\cite{Disalvo1976}, our calculations show that the CDW in MoS\s2 is driven by three nonequivalent M phonon instabilities and results in a \cdw2 unit cell with triangular distortions of Mo atoms [see Figs.~\ref{fig:fig1} and \ref{fig:fig2}(a) as well as Supplementary Figure~3], in agreement with the CDW pattern found in the recent STM measurements~\cite{BinSubhan2021}.
Due to the splitting of the bands [see Fig.~\ref{fig:fig3}(a)], the DOS at the Fermi level in a \cdw2 CDW phase drops, making it energetically favorable.
However, due to the existence of the two minima in the potential energy surface [see Fig.~\ref{fig:fig3}(c)], the atom relaxation in a \cdw2 supercell results in a hysteresis loop as doping is varied [see Fig.~\ref{fig:fig3}(d)].

\begin{figure}[!t]
    \centering
    \includegraphics[width=0.95\columnwidth]{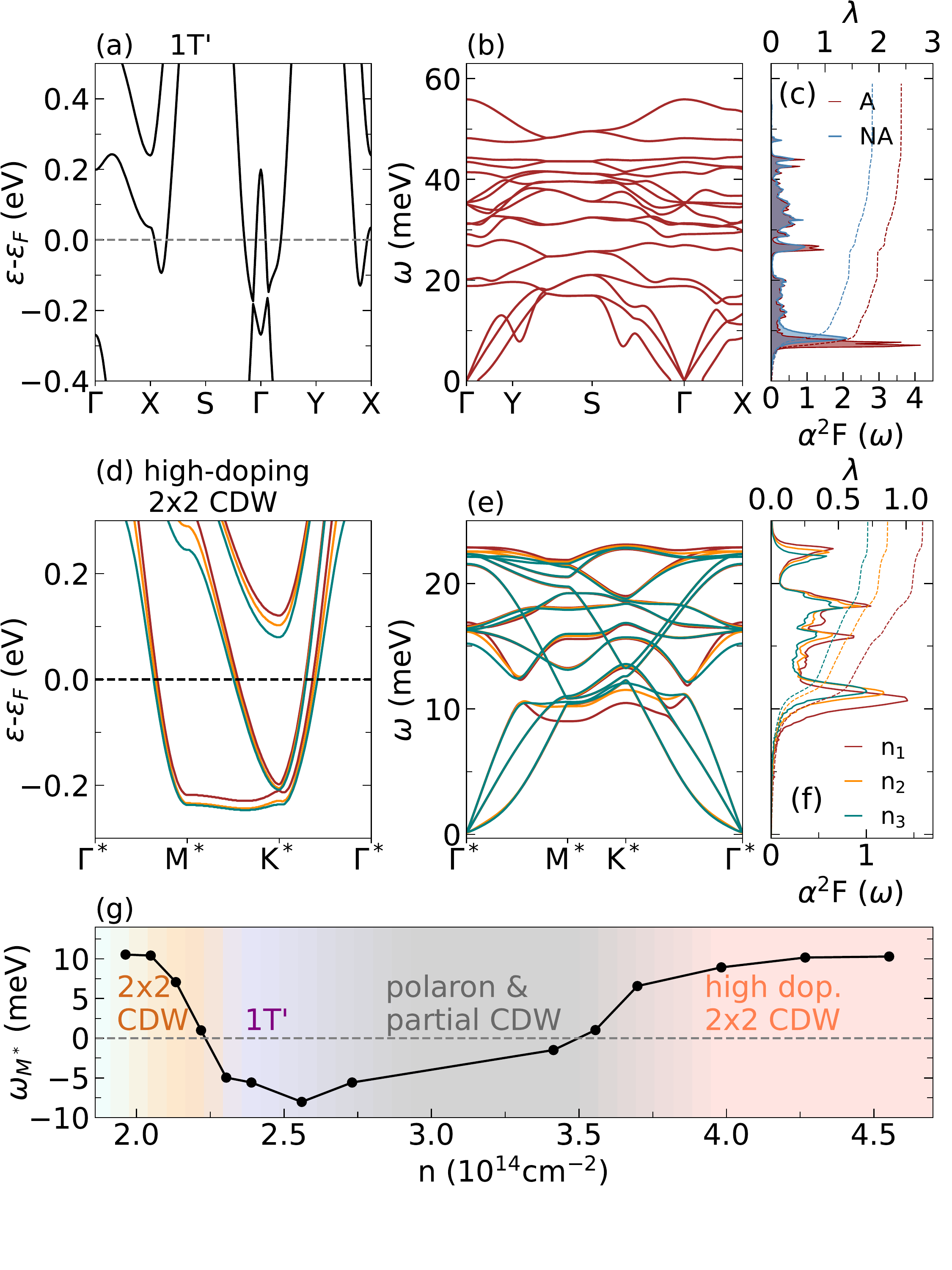}
    \caption{\textbf{Properties of MoS\s2 in the 1T$'$ and high-doping \cdw2 CDW phases.}
    (a)~Electron band structure obtained for the excess charge of $n = 2.62 \times 10^{14}\,\mathrm{cm}^{-2}$.
    (b)~Adiabatic phonon dispersion.
    (c)~Eliashberg spectral function calculated using the adiabatic (A) and nonadiabatic (NA) formula, with the corresponding cumulative EPC constant $\lambda(\omega)$.
    (d)~Electron band structures for the largest doping values considered in this work; $n_1 = 3.0 \times 10^{14}\,\mathrm{cm}^{-2}$, $n_2 = 4.3 \times 10^{14}\,\mathrm{cm}^{-2}$, $n_3 = 4.6 \times 10^{14}\,\mathrm{cm}^{-2}$.
    (e)~Corresponding phonon dispersions and (f)~NA Eliashberg spectral functions with cumulative EPC constants $\lambda(\omega)$.
    (g)~Softening and stabilization of the M$^\ast$ point acoustic phonon.
    The stabilization of this mode defines the low- and high-doping \cdw2 CDW phase, while the region where it is unstable corresponds to the polaronic and partial CDW phases.}
    \label{fig:fig4a}
\end{figure}

For the same critical doping value, we find that the \cdw1 H structure becomes unstable due to the aforementioned M point phonon instability, while a \cdw2 CDW structure becomes unstable due to the M$^\ast$ point (see Fig.~\ref{fig:fig1}) phonon instability.
The latter is clearly visible in Fig.~\ref{fig:fig3}(e).
In the \cdw1 H phase the M point instability is a result of electronic transitions between the filled K and 1/2\,K electron pockets on the Fermi surface.
In the \cdw2 CDW phase, band splitting leads to the occurrence of the flat band close to the Fermi level at the M point.
Electron transitions between these flat-band parts cause phonon softening in the new backfolded M$^\ast$ point (see Supplementary Figure~4).

\textbf{The 1T$'$ phase.}
Next, we find that one possible stable structure for larger doping values is a 1T$'$ phase, which is separated from the \cdw1 H phase by a large energy barrier of around 10\,eV.
This could explain why the transition to the 1T$'$ phase in experiments is not observed when MoS\s2 is doped by gating~\cite{Jin2018,Gan2018,Friedman2017}.
The 1T$'$ phase is a metastable phase, and our calculations confirm that the barrier reduces by doping~\cite{Huang2018} (see Supplementary Figure~5).
In Fig.~\ref{fig:fig4a}(a) we show the electronic bands and find that the 1T$'$ phase is stable for a narrow doping region which could be a consequence of its Fermi surface, which is prone to nesting.
The phonon spectral function and dispersion in Fig.~\ref{fig:fig4a}(c) reveal many Kohn anomalies in the lowest acoustic branch, appearing for the wave vectors which nest the Fermi surface.
Again, we observe large NA effects in the EPC properties [see Fig.~\ref{fig:fig4a}(d)].

\textbf{Polaronic distortions and partial CDW.}
Remarkably, after the \cdw1 H structure reaches instability, it does not immediately stabilize into the \cdw2 CDW phase.
As the \cdw1 M point backfolds in the \cdw2 $\Gamma$ point, a \cdw2 CDW phase seems to stabilize this particular instability [see Fig.~\ref{fig:fig3}(e)] but, as demonstrated above, only for the narrow doping region from 2.11 to $2.38 \times 10^{14}\,\mathrm{cm}^{-2}$.
By adding more electrons, this \cdw2 CDW phase again shows an instability in the new M$^\ast$ point (see Supplementary Figure~4), suggesting an existence of structural distortions on a larger length scale.

Calculations with supercells larger than \cdw2 are expensive within DFT and, therefore, we use a TBA model to investigate the region of our phase diagram where neither H nor \cdw2 CDW phases are stable (see grey area in Fig.~\ref{fig:fig1}).
Our model calculations show that stable structures indeed require much larger supercells, once the doping is increased past the first instability point.
Considering an \cdw{$18 \sqrt 3$} supercell, we find a single localized distortion, a polaron, where the triangular displacement of the three Mo atoms resembles that of a single unit of the \cdw2 CDW in Fig.~\ref{fig:fig2}(a).
With increasing doping, the number of localized distortions increases, forming first a partial and then a full \cdw{$3 \sqrt 3$} CDW.

\textbf{The high-doping \cdw2 CDW phase.}
At high doping, our DFT calculations show that a \cdw2 CDW phase once again becomes stable.
The soft acoustic phonon at the M$^\ast$ point of the \cdw2 BZ visible in Fig.~\ref{fig:fig3}(b) starts to blueshift as a function of doping, and becomes stable for around $n = 3.5 \times 10^{14}\,\mathrm{cm}^{-2}$, as depicted in Fig.~\ref{fig:fig4a}(g).
The flattening of the lowest conduction band, which hosts the excess charge [see Fig.~\ref{fig:fig4a}(d)], significantly lowers the DOS at the Fermi level.
As doping is increased, the M$^\ast$ point Kohn anomaly hardens [see Figs.~\ref{fig:fig4a}(e) and \ref{fig:fig4a}(g)], while the EPC constant reduces.

For this doping range, we also report a \cdw{$2 \sqrt 3$} CDW phase coexistence (see Supplementary Figure~6), in agreement with the experiment~\cite{BinSubhan2021}.
Here again the triangular pattern of Mo displacements is present, but in a modulated configuration where one in three triangles is more pronounced.
Using the TBA model, both a perfect \cdw2 CDW and massively distorted \cdw{$3 \sqrt 3$} CDWs are stable.

\begin{figure}[!t]
    \centering
    \includegraphics[width=\columnwidth]{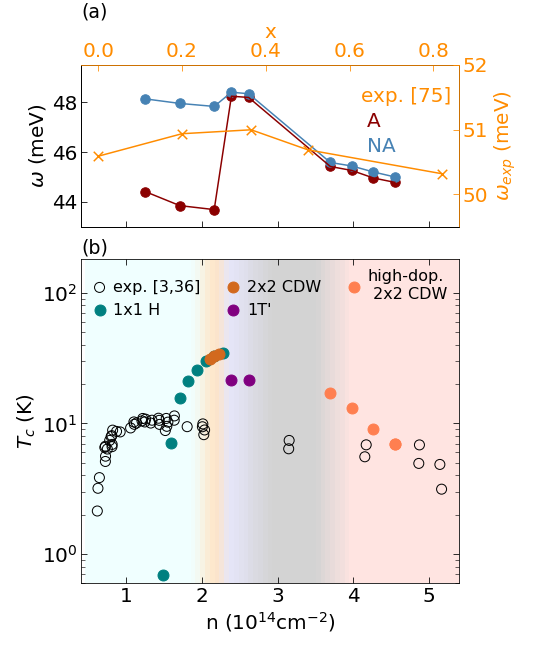}
    \caption{\textbf{Recreating the superconductivity dome structure of $T_c$ in doped MoS\s2.}
    (a)~Evolution of the A\s1 phonon mode frequency across the phase transition from the \cdw1 H to the 1T$'$ phase in the adiabatic and nonadiabatic approximation (lower x-axis).
    First-principles data is compared with the experiment from Ref.~\cite{Tan2018}, and the upper x-axis that corresponds to the experimental data is the atomic ratio of Li($x$) in Li$_x$MoS\s2.
    (b)~Doping-dependent SC phase diagram of MoS\s2, with different phases color-coded in the background.
    The experimental results for $T_c$ are presented with open black symbols~\cite{Ye2012,Costanzo2018}.}
    \label{fig:phase_diag}
\end{figure}

\begin{figure*}[!t]
    \centering
    \includegraphics[width=\linewidth]{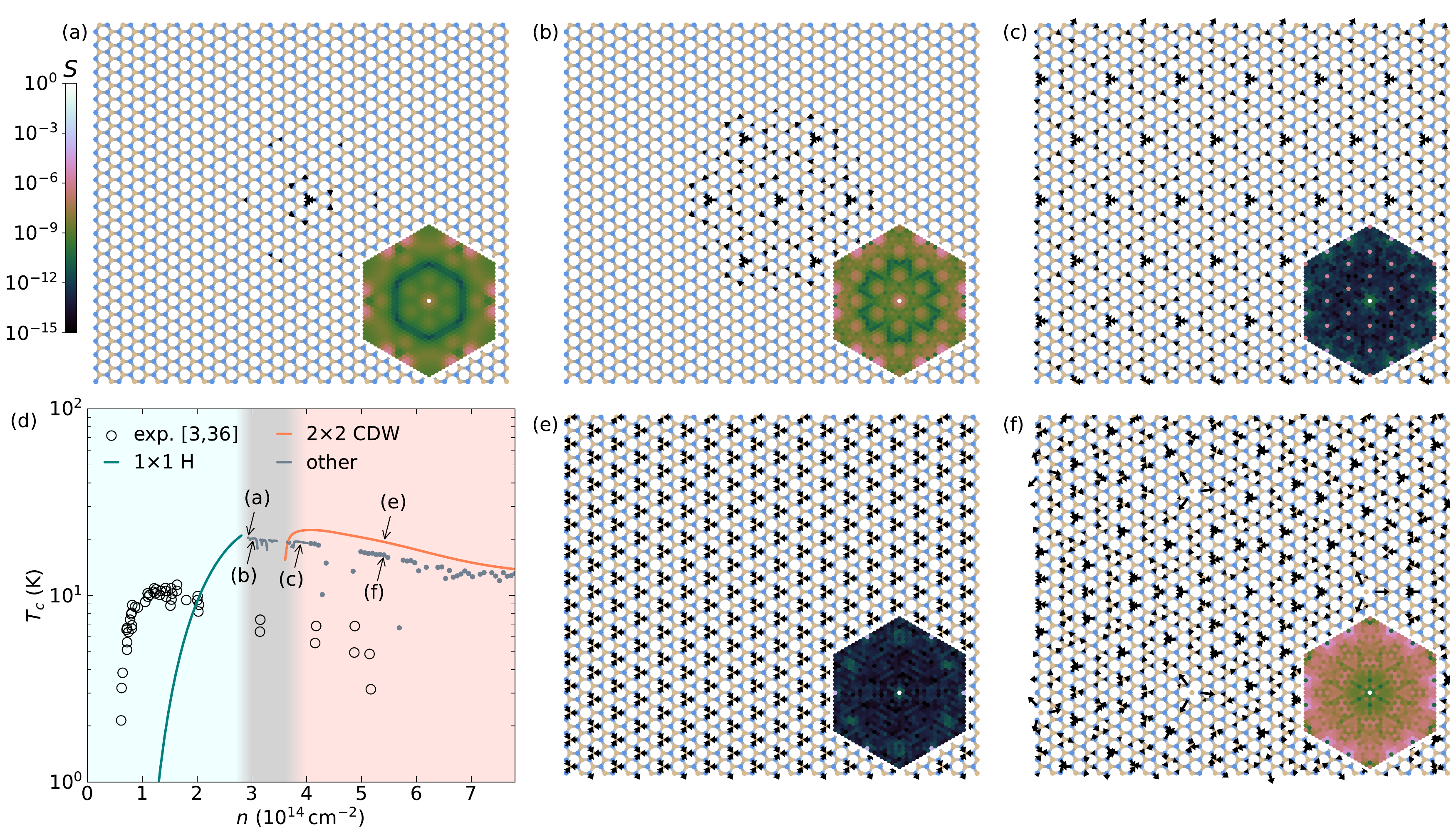}
    \caption{\textbf{First-principles-based tight-binding model for MoS\s2 monolayer.}
    All calculations have been performed on an \cdw{$18 \sqrt 3$} supercell, using the approximation of a linearized, nearest-neighbor EPC.
    Doping is modeled via a rigid shift of the chemical potential.
    \mbox{(a--c)}~Transition from polaronic deformation via partial to full \cdw{$3 \sqrt 3$} CDW for intermediate dopings.
    (d)~Critical temperature according to the Allen-Dynes formula as a function of doping for different structural phases.
    The underlying effective parameters $\lambda$ and $\omega_{\log}$ as well as the energy gain and displacement magnitude are shown in Supplementary Figure~7.
    The experimental reference points are the same as in Fig.~\ref{fig:phase_diag} and taken from Refs.~\cite{Ye2012,Costanzo2018}.
    (e,\,f)~\cdw2 CDW and strongly distorted phase at high doping.
    Arrows indicate atomic displacements, scaled by a factor of 20 for better visibility.
    Hexagonal insets show the structure factor $S=|\sum_{i = 1}^{N_{\mathrm{at}}} \exp(\mathrm i \vec q \vec R_i) / N_{\mathrm{at}}|^2$ for all supercell $\Gamma$ points $\vec q$ inside the first Brillouin zone of the primitive \cdw1 cell, which indicates periodicities present in the shown relaxed displacement patterns.}
    \label{fig:phase_diag_model}
\end{figure*}

\section*{Discussion}\label{sec:discussion}

To further validate the obtained structural phase diagram as a function of doping, in Fig.~\ref{fig:phase_diag}(a) we compare the calculated phonon frequency of Raman-active A$_{1g}$ optical phonon mode across various phases with the experimental measurements done on Li-doped (i.e., electron-doped) MoS\s2~\cite{Tan2018}.
This A$_{1g}$ optical phonon is an out-of-plane mode at the center of the BZ that is strongly coupled to the excess charge carriers in the conduction valleys of MoS\s2, and in the \cdw1 H phase is redshifted as a function of conduction band population~\cite{Sohier2019,Novko2020}.
However, in Raman measurements, the frequency of the A$_{1g}$ mode is blueshifted with the maximum value around $x = 0.4$ (where $x$ is the ratio of Li atoms with respect to MoS\s2 unit), and then redshifted for $x > 0.5$.
Our theoretical results presented in Fig.~\ref{fig:phase_diag}(a) confirm that the phase transitions between \cdw1 H, 1T$^{'}$, and CDW phases need to be taken into account in order to qualitatively capture the observed doping-induced renormalization of A$_{1g}$ frequency.
Additionally, the large NA frequency renormalization of the A$_{1g}$ mode for the \cdw1 H phase is present~\cite{Novko2020,GarciaGoiricelaya2020}, and it is crucial for understanding the modifications of the phonon frequency across the whole phase diagram.

In Figs.~\ref{fig:phase_diag} and \ref{fig:phase_diag_model}, we show a complete phase diagram of MoS\s2 obtained from first principles and model calculations, respectively.
Firstly, a monotonically increasing $T_c$ with doping persists until the point of instability.
In the region of \cdw1 H and \cdw2 CDW phase coexistence, we report the $T_c$ to be almost the same in the two phases and it reaches its peak value of 34.6\,K.
As the system assumes the 1T$'$ phase, the SC $T_c$ is lower, in agreement with experiments~\cite{Zhang2016}.

When the \cdw2 CDW phase becomes stable again for larger doping concentrations, due to hardening of the strongly coupled mode with doping and decrease of the DOS at the Fermi level, SC $T_c$ is reduced further.
Note that the doping concentrations corresponding to the experimentally observed regions of a \cdw2 and \cdw{$2 \sqrt 3$} CDW phases~\cite{BinSubhan2021} overlap with the region where the SC $T_c$ is decreasing~\cite{Costanzo2018}.
We believe that the indirect transition from the \cdw1 H phase to a \cdw2 CDW (possibly via polaronic phases) is the leading explanation of the SC dome.
Good qualitative agreement with the experimental dome structure~\cite{Ye2012,Costanzo2018} further corroborates this claim.
We note also that the dome structure of $T_c$ is correlated to the softening of the M$^\ast$ phonon, as one can see from Fig.~\ref{fig:fig4a}(g).

The above results are confirmed by the TBA model calculations.
We emphasize that the purpose of our model is a qualitative understanding, and not to provide precise quantitative values of SC $T_c$.
Figure~\ref{fig:phase_diag_model}(d) shows $T_c$ as a function of doping for different stable structures on the \cdw{$18 \sqrt 3$} supercell (see Methods).
In agreement with the DFT results in Fig.~\ref{fig:phase_diag}, we find that the \cdw1 H and \cdw2 CDW structures are favored for smaller and larger dopings, respectively.
For intermediate dopings, however, both are unstable toward the formation of polaronic distortions and partial CDWs as shown in Fig.~\ref{fig:phase_diag_model}\mbox{(a--c)}.
The periodicities present in these different types of structural order can be seen in the structure factor $S$ shown in the insets.
A possible explanation is that the doping electrons tend to localize near triangular distortions involving three Mo atoms, which can only cover the full sample as in the \cdw2 CDW if enough excess charge is available.
Here the TBA model calculations confirm that the saturation (or even slight decrease) of the DOS at the Fermi level for the polaronic and CDW phases, as well as the hardening of the relevant phonon modes, lead to the reduction of SC $T_c$ (see Supplementary Figure~7).
Our previous studies of such distortions in the MoS\s2 monolayer using the same model approach have revealed associated dispersionless in-gap states~\cite{vanEfferen2023,vanEfferen2024}.
Recently, similar distortions and bands have also been predicted for photodoped MoS\s2 and other TMDs~\cite{Holtgrewe2024}.
Also, a recent study of bulk MoTe\s2 argued that the laser-induced transition from 2H to 1T$'$ phase goes through polaron formation~\cite{Kim2024}.
Our combined DFT and TBA simulations suggest that such polaron formation might be universal and present in equilibrium conditions, playing an important role in structural phase transitions and formation of the SC dome.

Note that even though we reproduce the dome structure of $T_c$ in accordance with the experiments, our values of $T_c$ are always larger than the measured ones.
For instance, the maximum values obtained in Refs.~\cite{Ye2012} and \cite{Costanzo2018} are $T_c=10.9$\,K and $T_c=11.6$\,K, respectively.
Here we obtain $T_c=29.9$\,K for $n = 2.04 \times 10^{14}\,\mathrm{cm}^{-2}$, while this value is additionally reduced to $T_c=21.6$\,K when SOC is included.
In recent theoretical work, for the same doping concentration of $n = 2.02 \times 10^{14}\,\mathrm{cm}^{-2}$, a $T_c=19$\,K was reported~\cite{Marini2022}, and the importance of using a realistic gating geometry compared to homogeneous jellium background for reaching the more quantitative $T_c$ was discussed.
In Ref.~\cite{Das2015}, the renormalization of electronic states due to electron correlations was discussed as an important ingredient to obtain the quantitative agreement with the experimental $T_c$.
It should be also noted that the onset and the maximum value of the $T_c$ in our calculations is shifted towards larger doping concentrations compared to the experiments.
For instance, the critical doping at which the $T_c$ is maximum and where the original \cdw1 H structure becomes unstable is $n_c = 2.1 \times 10^{14}\,\mathrm{cm}^{-2}$, while in the experiments this value is around $n_c = 1.5 \times 10^{14}\,\mathrm{cm}^{-2}$~\cite{Ye2012}.
Nevertheless, this could be corrected by using the lower value for electronic temperature in DFT calculations, see also Supplementary Figure~1.

There are many studies discussing EPC strengths and the corresponding $T_c$ for doped MoS\s2~\cite{Ge2013,Rosner2014,Das2015,Fu2017,GarciaGoiricelaya2019,Marini2022,Veld2023,Sohier2024}, however, the dome structure and the complete phase diagram up to higher doping concentration is rarely discussed, even though intriguing SC dome appears in many different materials, including different TMDs~\cite{Lu2018}.

In Refs.~\cite{Marini2022} and \cite{Sohier2024} only the saturation of $T_c$ as a function of doping was theoretically obtained, while it was argued that the SC dome structure observed in the experiments is due to extrinsic effects, such as charge localization for higher dopings, which leads to strong disorder and therefore to effective increase of the Coulomb electron-electron repulsion, which competes with the phonon-mediated pairing~\cite{Anderson1983}.
This was modeled phenomenologically by increasing the Coulomb repulsion parameter $\mu^*$ that enters the Allen-Dynes formula and Eliashberg equations from 0.1 to 0.15.
In Ref.~\cite{Das2015} it is claimed that an increase of electron correlations with doping is an intrinsic effect, which suppresses the EPC and, therefore, the $T_c$, producing a dome structure.
However, the latter studies neglect the structural phase transitions, and the corresponding modifications to the electronic structure that are clearly present in MoS\s2 for larger doping concentrations.
In another work, the SC dome structure appears due to the parametrization of McMillan's formula~\cite{Rosner2014}, where a certain combination of EPC $\lambda$ and $\omega_{\log}$ could result in the saturation and decrease of $T_c$.
On the other hand, here we show that the $T_c$ for the \cdw1 H phase is always increasing as a function of doping (see also results in Ref.~\cite{Marini2022}) both for the Allen-Dynes and Migdal-Eliashberg approaches.
In the same study, the CDW phase was predicted to exist in doped MoS\s2, but the authors only reported the \cdw[2]1 CDW pattern coming from one nonequivalent M phonon instability, while here we report on the full \cdw2 CDW phase induced by the three nonequivalent soft phonon modes.
In addition, we show that this \cdw2 CDW phase in MoS\s2 is not stabilized immediately after the \cdw1 H phase becomes unstable, but is preceded by the polaronic states and partial CDWs as well as possibly by the 1T$'$ phase.
Also, the EPC strength and $T_c$ was not evaluated beyond the \cdw1 H phase in Ref.~\cite{Rosner2014}.

Here we show that the SC dome structure in doped MoS\s2 is related to the strongly-coupled soft phonon modes responsible for the structural phase transitions, as Figs.~\ref{fig:fig3}(b), \ref{fig:fig3}(e), and \ref{fig:fig4a}(g) as well as the model data in Supplementary Figure~7 suggest.
Namely, when approaching the point of structural instability with doping, the relevant phonon mode becomes softer and therefore the EPC $\lambda$ and $T_c$ are increased.
With a further increase of doping, the new structure is stabilized, the relevant phonon mode hardens, and $\lambda$ and $T_c$ are decreasing, which in total creates the dome structure.
The importance of the soft phonon modes in elevating the $T_c$ was discussed a long time ago for A15 compounds~\cite{Testardi1972,Allen1974a}, and it was proposed as a source of the unconventional SC dome structure in SrTiO\s3~\cite{Appel1969,Edge2015} as well as predicted to play a role in BaTiO\s3~\cite{Ma2021} and WTe\s2~\cite{Yang2020}.
For SrTiO\s3, critical doping concentration for which the $T_c$ reaches maximum value correlates with the point when the optical polar phonon at the center of the BZ becomes unstable, inducing ferroelectric fluctuations~\cite{Edge2015}.
Here, instead of the ferroelectric phase, the maximal $T_c$ is related to the CDW and polaronic phase transitions and the corresponding soft acoustic phonon modes away from the BZ center.

In conclusion, we have explored the phase diagram of electron-doped MoS\s2 and explained the SC dome structure observed in various transport and tunneling experiments.
Across a large doping range, we have found a number of dynamically stable phases, such as the \cdw2 CDW, the 1T$'$ phase, polaronic states, and partial CDWs.
In agreement with existing literature, we have found a doping induced increase in $T_c$.
Following a steep increase, a combination of structural phase transitions and CDW formations leads to its reduction.
The latter correlates with the softening and hardening of the relevant acoustic phonon modes that are both responsible for the structural phase transitions and dominate the SC pairing mechanism.
Our work reconciles various experimental finding of CDWs in MoS\s2, with its doping-dependent SC dome structure, and could be helpful in comprehending the SC mechanism of other materials hosting a $T_c$ dome, such as other correlated TMDs and SrTiO\s3.

\section*{Methods}

\textbf{Migdal-Eliashberg theory.}
When it comes to calculating the $T_c$, there exist several levels of complexity, building up on the pioneering work of Bardeen, Cooper, and Schrieffer (BCS)~\cite{Bardeen1957}.
Specific details of a material's electronic structure, phonon spectrum, and interaction strengths can be included within the scope of Migdal-Eliashberg theory~\cite{Eliashberg1960}.
The smoothness of the Fermi surface determines whether an isotropic set of Eliashberg equations can be used~\cite{Marsiglio1992,Berges2016a}:
\begin{subequations}
\label{eq:eliashberg}
\begin{align}
    \widetilde \omega_n &= \omega_n + T
    \int_{-\infty}^\infty \mathrm d \varepsilon
    \frac{N(\varepsilon)}{N(\varepsilon_F^0)}
    \sum_m
    \frac{\widetilde \omega_m}{\Theta_m(\varepsilon)}
    \lambda_{n - m},
    \\
    \phi_n &= T
    \int_{-\infty}^\infty \mathrm d \varepsilon
    \frac{N(\varepsilon)}{N(\varepsilon_F^0)}
    \sum_m
    \frac{\phi_m}{\Theta_m(\varepsilon)}
    [\lambda_{n - m} - \mu],
    \label{eq:phi}
    \\
    \chi_n &= -T
    \int_{-\infty}^\infty \mathrm d \varepsilon
    \frac{N(\varepsilon)}{N(\varepsilon_F^0)}
    \sum_m
    \frac{\varepsilon - \varepsilon_F + \chi_m}{\Theta_m(\varepsilon)}
    \lambda_{n - m},
    \\
    n_e &=
    \int_{-\infty}^\infty \mathrm d \varepsilon
    N(\varepsilon)
    \biggl[
        1 - 2 T \sum_m
        \frac{\varepsilon - \varepsilon_F + \chi_m}{\Theta_m(\varepsilon)}
    \biggr],
\end{align}
with the common denominator
\begin{equation}
    \Theta_n(\varepsilon) = \widetilde \omega_n^2 + \phi_n^2 + [\varepsilon - \varepsilon_F + \chi_n]^2.
\end{equation}
\end{subequations}
Here, $N(\varepsilon)$ is the DOS per spin, $\omega_n = (2 n + 1) \pi T$ are fermionic Matsubara frequencies, $\widetilde \omega_n$ are renormalized Matsubara frequencies, $\phi_n$ is the SC order parameter, and $\chi_n$ an energy shift.
The number of electrons $n_e$ is conserved, which implies that the chemical potential $\varepsilon_F$ may deviate from its noninteracting value $\varepsilon_F^0$.
The Eliashberg equations are solved iteratively for $\widetilde \omega_n$, $\phi_n$, $\chi_n$, and $\varepsilon_F$, using our own implementation~\cite{Berges2016b}.
The presence of a SC state ($\phi_n > 0$) for a given temperature $T$ is determined by the effective Coulomb interaction $\mu$ and the effective EPC
\begin{equation}
    \lambda_n = \int_0^\infty \mathrm d \omega
    \frac{2 \omega}{\omega^2 + \nu_n^2}
    \alpha^2 F(\omega),
\end{equation}
where $\nu_n = 2 n \pi T$ are bosonic Matsubara frequencies.
Since we are only interested in the value of $T_c$, it is sufficient to only consider the linearized version ($\phi_n \rightarrow 0$) of the equations.

From the linearized form of the (real-axis) Eliashberg equations, McMillan's formula can be qualitatively obtained~\cite{McMillan1968}.
The majority of theoretical estimates of $T_c$ rely on McMillan's semiempirical result, which was later refined by Allen and Dynes~\cite{Allen1975}:
\begin{equation}
    T_c= \frac{f_1f_2\omega_{\log}}{1.2} \exp\left[-\frac{1.04(1+\lambda)}{\lambda-\mu^*-0.62\lambda\mu^*}\right].
    \label{eq:tc}
\end{equation}
The strong-coupling correction $f_1 = [1 + (\lambda / \Lambda_1)^{3 / 2}]^{1 / 3}$ and the shape correction $f_2 = 1 + (r - 1) \lambda^2 / (\lambda^2 + \Lambda_2^2)$ with $\Lambda_1 = 2.46 (1 + 3.8 \mu^*)$, $\Lambda_2 = 1.82 (1 + 6.3 \mu^*) r$, and $r = \overline \omega_2 / \omega_{\log}$ are needed if $\lambda > 1.5$.
We refer to Eq.~\eqref{eq:tc} as the Allen-Dynes formula, while the same result without the strong-coupling corrections is referred to as McMillan's formula.
EPC effects are encapsulated in the EPC constant $\lambda$ and the logarithmic and second-moment average phonon frequencies $\omega_{\log}$ and $\overline \omega_2$, which are obtained as integrals over the Eliashberg spectral function:
\begin{align}
    \lambda &= 2 \int^{\infty}_{0} \frac{\alpha^2F(\omega)}{\omega} \mathrm d\omega = \lambda_0, \label{eq:lambda}
    \\
    \omega_{\log} &= \exp\left[\frac{2}{\lambda}\int^{\infty}_{0} \log(\omega)\frac{\alpha^2F(\omega)}{\omega} \mathrm d\omega\right], \label{eq:omegalog}
    \\
    \overline \omega_2 &= \sqrt{\frac 2 \lambda \int_0^\infty \omega \alpha^2 F(\omega) \mathrm d \omega}. \label{eq:omegabar}
\end{align}

A third material parameter is the Coulomb pseudopotential.
The effective Coulomb electron-electron interaction $\mu$ in Migdal-Eliashberg theory differs from the Morel-Anderson pseudopotential $\mu^*$ of McMillan's formula~\cite{Morel1962}.
The latter is scaled to a typical phonon frequency, which we assume to be $\overline \omega_2$ following Ref.~\cite{Allen1975}.
For a given $\mu^*$, we can then estimate
\begin{equation}\label{eq:mu1}
   \mu = [1 / \mu^* - R(\overline \omega_2)]^{-1},
\end{equation}
with the cutoff-dependent rescaling term~\cite{Berges2016a}
\begin{equation}
    R(\omega) = \frac 1 \pi
    \int_{-\infty}^\infty \mathrm d \varepsilon
    \frac{N(\varepsilon)}{N(\varepsilon_F^0)}
    \frac 1 {\varepsilon - \varepsilon_F^0}
    \arctan \left[ \frac{\varepsilon - \varepsilon_F^0}{\omega} \right].
\end{equation}
Also the fact that the summations over Matsubara frequencies are in practice truncated at a cutoff frequency $\omega_C$ has to be considered.
We limit the summations in Eqs.~\eqref{eq:eliashberg} to $|\omega_m| < \omega_C$ and at the same time replace $\mu$ in Eq.~\eqref{eq:phi} by $\mu_{\rm C}^* = [1 / \mu + R(\omega_C)]^{-1}$ scaled to the Matsubara cutoff.
Using Eq.~\eqref{eq:mu1}, we can also write
\begin{equation}
    \mu_{\rm C}^* = [1 / \mu^* - R(\overline \omega_2) + R(\omega_C)]^{-1},
\end{equation}
where $\mu^*$ depends on doping and in multi-valley materials it also changes when calculating the inter/intra-valley repulsion.
For larger doping values it has been calculated to have the mean value of 0.13~\cite{Schonhoff2016}.
In this work we therefore set $\mu^*=0.13$ and use it as a constant.

The Eliashberg spectral function in Eqs.~\eqref{eq:lambda}--\eqref{eq:omegabar} can be written as~\cite{Allen1974b,Allen1982}
\begin{equation}
    \alpha^2F(\omega) = \frac{1}{\pi N(\varepsilon_F^0)}\sum_{\mathbf{q}\nu}\frac{\gamma_{\mathbf{q}\nu}}{\omega_{\mathbf{q}\nu}}B_{\nu}(\mathbf{q},\omega),
    \label{eq:a2f}
\end{equation}
where $\mathbf{q}$ and $\nu$ are phonon wavevector and band indices.
$\omega_{\mathbf{q}\nu}$ are the statically screened phonon frequencies, as obtained in density functional perturbation theory (DFPT)~\cite{Baroni2001,Giannozzi2009}.
In Eq.~\eqref{eq:a2f}, $\alpha^2F(\omega)$ is written in its NA form.
It can be seen as a phonon DOS, weighted, renormalized, and broadened by EPC in the form of phonon linewidths $\gamma_{\mathbf{q}\nu}$ obtained in the double-delta approximation~\cite{Allen1972} and the NA phonon spectral function
\begin{equation}
    B_{\nu}(\mathbf{q},\omega)=-\frac{1}{\pi}\Im\left[ \frac{2\omega_{\mathbf{q}\nu}}{\omega^2-\omega_{\mathbf{q}\nu}^2-2\omega_{\mathbf{q}\nu}\pi_{\nu}(\mathbf{q},\omega)} \right].
    \label{eq:phon_spec}
\end{equation}
With the term `adiabatic', we refer to the statically screened and infinitely long lived (DFPT) phonons~\cite{Marini2024}.
Nonadiabaticity refers to the fact that the statically screened DFPT phonon frequencies are corrected by the real part of the NA phonon self-energy
\begin{multline}
    \pi_{\nu}(\mathbf{q},\omega)=\sum_{\mathbf{k}nm}\left| g_{mn\nu}(\mathbf{k},\mathbf{q}) \right|^2\frac{f_{n\mathbf{k}}-f_{m\mathbf{k+q}}}{\omega+\varepsilon_{n\mathbf{k}}-\varepsilon_{m\mathbf{k+q}}+i\eta}\\ -\sum_{\mathbf{k}nm}\left| g_{mn\nu}(\mathbf{k},\mathbf{q}) \right|^2\frac{f_{n\mathbf{k}}-f_{m\mathbf{k+q}}}{\varepsilon_{n\mathbf{k}}-\varepsilon_{m\mathbf{k+q}}}
    \label{eq:pi_na}
\end{multline}
evaluated at the unrenormalized frequencies, including the broadening effects as well.
In Eq.~\eqref{eq:pi_na}, $\mathbf{q}\neq 0$, $\mathbf{k}$ and $n,m$ are electronic momentum and band indices, $f_{n\mathbf{k}}$ denote the Fermi-Dirac distribution functions, $\varepsilon_{m\mathbf{k+q}}$ are the electron energies, and $g_{mn\nu}(\mathbf{k},\mathbf{q})$ are the statically screened EPC matrix elements.
We use the `screened-screened' approximation to the phonon self-energy for quadratic errors~\cite{Berges2023}.

\textbf{Model calculations.}
To estimate the critical temperature for lattice distortions on very large supercells (here 2916 atoms), for which full DFPT calculations are unfeasible, we use the approximation of a linearized EPC as described in Ref.~\cite{Schobert2024}, wherein it is referred to as `model III'.
Additionally, we reduce the EPC to a nearest-neighbor model including only the $d_{z^2}$, $d_{x^2 - y^2}$, and $d_{x y}$ orbitals of the transition-metal atoms, as defined in Eq.~(B4) of Ref.~\cite{Berges2020} for TaS\s2, with parameters fitted to DFPT data for MoS\s2.
The first-principles data are taken from Appendix~B of Ref.~\cite{Berges2023}.
Doping is modeled as a rigid shift of the chemical potential.
The electronic temperature is set to $5\,\mathrm{mRy} \approx 800\,\mathrm K$.
For each carrier concentration the atomic positions are optimized, starting from both a randomly distorted \cdw2 CDW and a single triangular displacement involving three neighboring Mo atoms, until all forces are below $10^{-5}$\,Ry/bohr.
Phonons and the adiabatic effective interaction parameters are then calculated for $\mathbf{q} = 0$ only, with the help of sparse matrices and repeated transforms between the orbital and band basis to save memory~\cite{Berges2017}.

\textbf{Computational details.}
The written equations and their ingredients are calculated by means of DFPT~\cite{Baroni2001} and Wannier interpolation~\cite{Mostofi2008} of EPC matrix elements~\cite{Ponce2016}.
We use \textsc{Quantum ESPRESSO}~\cite{Giannozzi2009} for the DFT calculations, and the EPW code~\cite{Giustino2007,Noffsinger2010,Ponce2016,Lee2023} for EPC.
We use the fully-relativistic norm-conserving Perdew-Burke-Ernzerhof pseudopotentials with a kinetic energy cutoff of 100 Ry from \textsc{PseudoDojo}~\cite{VanSetten2018}.
The NA dynamic and static term in the phonon self-energy, Eq.~\eqref{eq:pi_na}, are calculated separately, because in the dynamic term, dense $\mathbf{k}$- and $\mathbf{q}$-grids are used, while the grids for the static-term calculation are already determined by the DFT and DFPT calculations~\cite{GarciaGoiricelaya2019,Novko2018,Girotto2023}.
For the calculations in a \cdw1 H phase, the relaxed lattice constant is set to 3.186~\AA, with vacuum of 16~\AA.
The self-consistent electron density calculation is done on a $24 \times 24 \times 1$ $\mathbf{k}$-point grid and the phonon calculation on a $12 \times 12 \times 1$ $\mathbf{q}$-point grid.
In EPW, we use 11 maximally localized Wannier functions~\cite{Marzari2012} with the initial projections of $d$~orbitals on the Mo sites and $p$~orbitals on the S atom sites.
The fine sampling of the Brillouin zone for the electron-phonon interpolation is done on a $360 \times 360 \times 1$ $\mathbf{k}$-grid and $120 \times 120 \times 1$ $\mathbf{q}$-grid.
Smearing in the EPW calculation is set to 20\,meV.
For the calculations in other phases, all the abovementioned parameters are the same, except that the unit cell is larger and the grids are accordingly smaller.
In all the DFT and DFPT calculations we use the Coulomb truncation~\cite{Sohier2017}, while in the EPW calculation we use the long-range terms which include 2D dipoles following Ref.~\cite{Ponce2023}.

%

\bibliography{ref}

\begin{acknowledgments}
Useful discussions with Raffaello Bianco are gratefully acknowledged. N.\,G.\,E. and D.\,N. acknowledge financial support from the Croatian Science Foundation (Grant no. UIP-2019-04-6869).
J.\,B. acknowledges support from the Deutsche Forschungsgemeinschaft (DFG) under Germany's Excellence Strategy (University Allowance, EXC~2077, University of Bremen), computing time granted by the Resource Allocation Board and provided on the supercomputer Emmy/Grete at NHR-Nord@G\"ottingen as part of the NHR infrastructure, and fruitful discussions with Tim O. Wehling, Arne Schobert, and Laura P\"atzold.
S.\,P. acknowledges support from the Fonds de la Recherche Scientifique de Belgique (FRS-FNRS) and by the Walloon Region in the strategic axe FRFS-WEL-T.
Computational resources have been provided by the PRACE award granting access to MareNostrum4 at Barcelona Supercomputing Center (BSC), Spain and Discoverer in SofiaTech, Bulgaria (OptoSpin project id.~2020225411).
\end{acknowledgments}


%
%
%
%



%

\end{document}